\documentstyle[twoside,fleqn,espcrc2,epsf]{article}

\newcommand{\beq}{\begin{equation}}
\newcommand{\eeq}{\end{equation}}
\newcommand{\beqa}{\begin{eqnarray}}
\newcommand{\eeqa}{\end{eqnarray}}

\def\spose#1{\hbox to 0pt{#1\hss}}
\def\ltapprox{\mathrel{\spose{\lower 3pt\hbox{$\mathchar"218$}}
 \raise 2.0pt\hbox{$\mathchar"13C$}}}
\def\gtapprox{\mathrel{\spose{\lower 3pt\hbox{$\mathchar"218$}}
 \raise 2.0pt\hbox{$\mathchar"13E$}}}
\def\inapprox{\mathrel{\spose{\lower 3pt\hbox{$\mathchar"218$}}
 \raise 2.0pt\hbox{$\mathchar"232$}}}

\newcommand{\AmS}{{\protect\the\textfont2
  A\kern-.1667em\lower.5ex\hbox{M}\kern-.125emS}}

\title{QCD with dynamical Wilson fermions at $\beta=5.5$\thanks{Preprint
FSU-SCRI-96-59. To appear in the proceedings of Lattice '96, St. Louis,
Missouri, USA, 4--8 June. Presented by U.~M.~Heller.}}

\author{K.M.~Bitar, R.G.~Edwards, U.M.~Heller, and
        A.D.~Kennedy\address{SCRI, The Florida State University,
                             Tallahassee, FL 32306-4052, USA}}

\begin{document}

\begin{abstract}
We study QCD with two flavors of dynamical Wilson fermions at $\beta = 5.5$
and three values of $\kappa$. The corresponding pion masses are 0.375,
0.324 and 0.262 in lattice units, with pion to rho mass ratios of 0.76,
0.71 and 0.62, respectively. We use the configurations to compute the heavy
quark potential, leading to lattice spacings of 0.110, 0.105 and 0.099 fm,
and to compute spectroscopy for several different valence quark $\kappa$'s.
\end{abstract}

\maketitle


\section{INTRODUCTION}
\label{introduction}

Of the two popular unimproved ways of discretizing the Dirac operator simulations with
dynamical Wilson fermions still lag behind simulations with dynamical
staggered fermions. The best completed simulation with Wilson fermions, by
the HEMCGC collaboration \cite{HEMCGC_W94}, was done at stronger couplings,
{\it i.e.,} larger lattice spacing, than the state-of-the-art staggered
simulations \cite{Columbia91_95,HEMCGC_S94}, and at larger $m_\pi/m_\rho$
ratios, 0.72 and 0.60. The lowest $m_\pi/m_\rho$ ratio achieved to date
with staggered fermions, in contrast, is 0.33 \cite{Columbia94}, albeit at
quite a strong coupling. There are many indications that the Wilson fermion
HEMCGC
simulations were done far from the continuum limit: for example $a^{-1}$,
in the chiral limit, was estimated as 1600 MeV from the nucleon mass, 1800
MeV from the rho mass and 2000 MeV from $f_\pi$, and a computation of
$\Delta \beta$ proved impossible \cite{beta_fn}.

\begin{figure}
\begin{center}
\vskip 10mm
\leavevmode
\epsfxsize=60mm
\epsfbox[80 80 550 410]{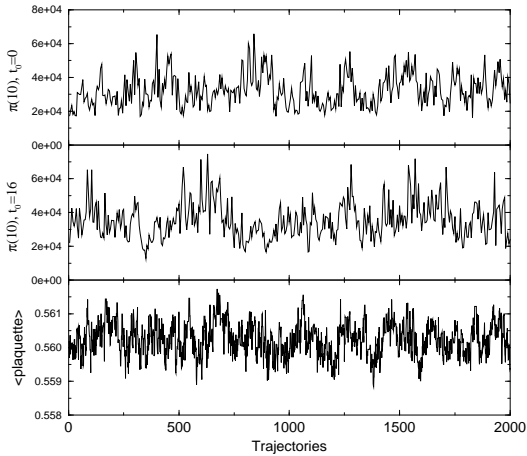}
\vskip -5mm
\end{center}
\caption{Time history of the plaquette and the pion correlation function
         at distance 10 with wall source at $t_0=0$ and 10 for
         $\kappa=0.1600$.}
\label{fig:hist_1600}
\vskip -5mm
\end{figure}

About 3 years ago, the SCRI lattice group decided to push dynamical Wilson
fermion simulations to a somewhat weaker coupling with runs on the SCRI
CM-2. Being limited, for technical reasons, to lattice sizes that are a
power of two, we chose a $16^3 \times 32$ lattice and a gauge coupling
$\beta=5.5$. With these choices we hoped that the finite size effects will
be relatively small. We have made runs at three different $\kappa$ values,
to allow chiral fits and thus to test the chiral extrapolations. In each
case we have about 2000 trajectories, after thermalization, with parameters
given in Table~\ref{tab:params}. Table~\ref{tab:results} contains
some results, including rough estimates of
integrated autocorrelation times.
Sample time histories, for the run with $\kappa=0.1600$, are shown in
Fig.~\ref{fig:hist_1600}.

\begin{table}
  \begin{center}
    \caption{Summary of run parameters}
    \label{tab:params}
    \tabcolsep 4pt
    \vspace{2mm}
    \begin{tabular}{|c|cccc|}
      \hline
      $\kappa$ & $dt$ & $R$ & Acc & $\langle N_{CG} \rangle$ \\ \hline
      0.1596 & 0.01754 & $3\cdot10^{-7}$ & 75\% & 270 \\
      0.1600 & 0.01515 & $3\cdot10^{-7}$ & 75\% & 380 \\
      0.1604 & 0.007   & $3\cdot10^{-7}$ & 90\% & 450 \\ \hline
    \end{tabular}
  \end{center}
  \vskip -10mm
\end{table}

\begin{table}
  \begin{center}
    \caption{Some results}
    \label{tab:results}
    \tabcolsep 4pt
    \vspace{2mm}
    \begin{tabular}{|c|cccc|}
      \hline
      $\kappa$ & $\langle \frac{1}{3}{\rm Tr}U_p \rangle$ &
               $\alpha_V(\frac{3.41}{a})$ & 
               $\tau_{\rm plaq}$ & $\tau_{\pi(10)}$ \\ \hline
      0.1596 & 0.55928(9) & 0.178 & 33 & 65 \\
      0.1600 & 0.56025(5) & 0.177 & 9  & 15 \\
      0.1604 & 0.56131(8) & 0.176 & 21 & 21 \\ \hline
    \end{tabular}
  \end{center}
  \vskip -10mm
\end{table}

\section{THE HEAVY QUARK POTENTIAL}
\label{potential}

On the ensembles of configurations generated, we computed the heavy quark
potential from a combination of on- and off-axis timelike Wilson loops, with
``APE smeared" links in the space directions \cite{APE87} to enhance the
signal to noise ratio. Since string breaking is estimated to occur at a
distance somewhat larger than accessible to us (and we do not observe any
sign of it)  we used the same ansatz that is common in quenched simulations
for fitting the potential
\beq
  V(\vec{r}) = V_0 + \sigma r
    - {e\over r} - e'\Bigl(G_L(\vec{r}) - {1\over r}\Bigr);
  \label{eq:fit_form}
\eeq
with $G_L$, the lattice Coulomb potential, taking account of the lattice
artefacts present at small distances. We used fully correlated fits with
the covariance matrix estimated by a jackknife method.

\begin{figure}
\begin{center}
\vskip -10mm
\leavevmode
\epsfxsize=60mm
\epsfbox[120 80 530 500]{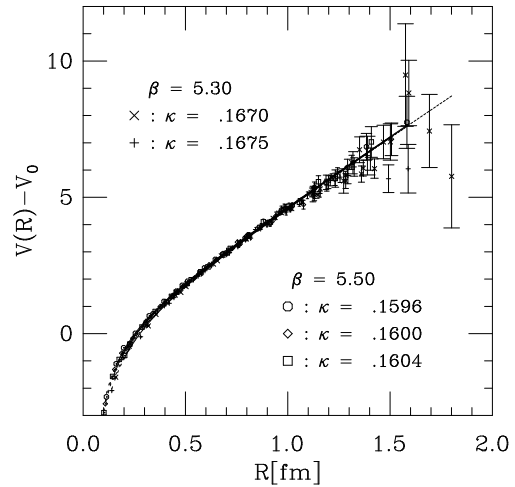}
\vskip -3mm
\end{center}
\caption{The potential, in physical units, including, for comparison,
         the $\beta=5.3$ data from \protect\cite{qcd_pot}.}
\label{fig:resc_pot}
\vskip -5mm
\end{figure}

The parameters for the best fits, defined as in ref.~\cite{qcd_pot}, are
listed in Table~\ref{tab:fit_jack}. We show the potentials together with
the fits, in physical units as determined from the string tension, in
Fig.~\ref{fig:resc_pot}. Also included in the plot are the potentials from
the $\beta=5.3$ simulations as computed in ref.~\cite{qcd_pot}. All curves
agree fairly well, showing approximate scaling of the potential. The
Coulomb coefficient $e$ is somewhat larger than for quenched calculations
at comparable lattice spacing, which can be attributed to the effect of the
dynamical fermions \cite{SESAM_pot}.

\begin{table*}
  \begin{center}
    \caption{Summary of results from fits to the effective potentials using
             eq.~(\protect\ref{eq:fit_form}). The last two columns give the
             scale $r_0/a$ determined from $r_0^2 F(r_0) = 1.65$ and the
             dimensionless quantity $r_0 \protect\sqrt{\sigma}$.}
    \label{tab:fit_jack}
    \tabcolsep 4pt
    \vspace{2mm}
    \begin{tabular}{|l|l|lr|cccc|c|cc|}
      \hline
      $\kappa$ & $T$ & $r_{\rm min}$ & $r_{\rm max}$ &
        $a V_0$ & $a^2 \sigma$ & $e$ & $e'$ &
        $Q$ & $r_0/a$ & $r_0 \sqrt{\sigma}$ \\ \hline
      0.1596 & 4 & 1.41 & 13.86 & 0.731(3) & 0.0645(7) & 0.315(3) & 0.368(12) &
        0.90 & 4.55(2) & 1.155(1) \\ \hline
      0.1600 & 4 & 1.73 & 13.86 & 0.743(5) & 0.0583(9) & 0.329(6) & 0.375(13) &
        0.77 & 4.76(3) & 1.149(3) \\ \hline
      0.1604 & 4 & 2.24 & 13.86 & 0.757(6) & 0.0510(9) & 0.342(5) & 0.53(8) &
        0.81 & 5.06(3) & 1.144(4) \\ \hline
    \end{tabular}
  \end{center}
\end{table*}

\section{SPECTROSCOPY}
\label{spectroscopy}

\begin{figure}
\begin{center}
\vskip -5mm
\leavevmode
\epsfxsize=60mm
\epsfbox[150 160 530 490]{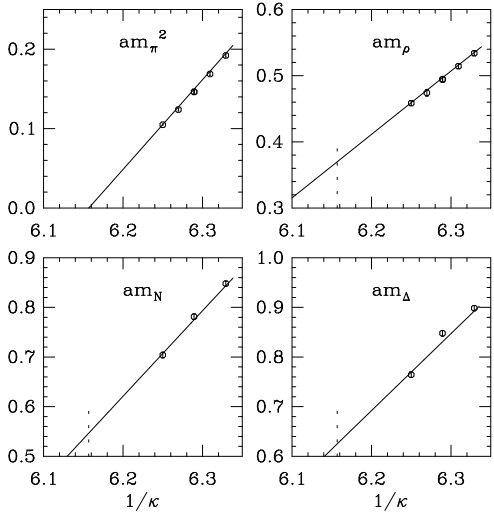}
\end{center}
\caption{Chiral plots for the hadron spectroscopy on the ensemble with
         $\kappa_{\rm sea}=0.1600$. The dashed lines indicate $\kappa_c$.}
\label{fig:chir_1600}
\vskip -5mm
\end{figure}

For each ensemble we computed the hadron spectroscopy with six different
valence quark kappas, $\kappa_V = \kappa_{\rm sea}$, 0.1590, 0.1580,
0.1395, 0.1230 and 0.1050. The last three are quite heavy, in the $c$ to
$b$ quark range, and will not be discussed here. Mesons were computed for
all pairs of $\kappa_V$'s, while baryons only for equal valence quarks.
Mass fits, preferably to two states, and chiral fits were fully correlated.
In Fig.~\ref{fig:chir_1600} we show chiral plots (vs. $1/\kappa$) for the
results with $\kappa_{\rm sea}=0.1600$. Similar plots for the other sea
quarks and tables of all the masses can be found in ref.~\cite{beta_5p5}.

\begin{figure}
\begin{center}
\leavevmode
\epsfxsize=60mm
\epsfbox[150 160 530 490]{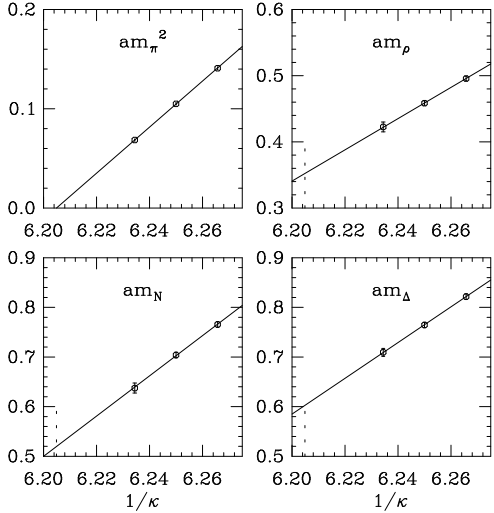}
\end{center}
\caption{Chiral plots for the hadron spectroscopy with $\kappa_V =
         \kappa_{\rm sea}$.}
\label{fig:chir_5p5}
\vskip -5mm
\end{figure}

Having ensembles with three different sea quark masses, we can also make
chiral fits for spectroscopy involving only $\kappa_V = \kappa_{\rm sea}$.
These are shown in Fig.~\ref{fig:chir_5p5}. We obtain $\kappa_c =
0.16116(15)$ and extrapolated masses $am_\rho = 0.352(12)$, $am_N =
0.520(18)$ and $am_\Delta = 0.603(15)$. These numbers, as well as the
results in Table~\ref{tab:a_MeV} for $\kappa_{\rm sea} = 0.1604$, are still
preliminary. We are in the process of computing propagators from another
time slice.

In Table~\ref{tab:a_MeV} we give a collection of estimates of $a^{-1}$ in
MeV, obtained from the potential, rho, nucleon and Delta masses. For the
latter, at fixed $\kappa_{\rm sea}$, the masses were extrapolated to the
physical $m_\pi/m_\rho$ ratio. The last line shows the extrapolation to
zero sea quark mass from the masses obtained with $\kappa_V = \kappa_{\rm
sea}$. As can be seen, the lattice spacing estimates still differ
considerably across each line in the table and we have to conclude that
there are still rather large O($a$) effects present. It will be difficult
to decrease these lattice effects by simply going to smaller lattice
spacings.

\begin{table}
  \begin{center}
    \caption{Values for $a^{-1}$ in MeV. The last line comes from
             extrapolations in the sea quark mass to the chiral limit.} 
    \label{tab:a_MeV}
    \tabcolsep 4pt
    \vspace{2mm}
    \begin{tabular}{|c|cccc|}
      \hline
      $\kappa$ & $r_0$ & $\rho$ & $N$ & $\Delta$ \\ \hline
      0.1596  & 1790(10) & 1955(40) & 1650(30) & 1910(45) \\
      0.1600  & 1885(15) & 2025(25) & 1685(30) & 1950(45) \\
      0.1604  & 1995(15) & 1995(40) & 1755(45) & 1985(40) \\ \hline
      0.16116 & 2185(25) & 2190(75) & 1805(65) & 2045(55) \\ \hline
    \end{tabular}
  \end{center}
  \vskip -10mm
\end{table}

\section*{Acknowledgements}

This work was partly supported by the DOE under grants
\#~DE-FG05-85ER250000 and \#~DE-FG05-92ER40742. The computations were
carried out on the CM-2 at SCRI.


\end{document}